\pdfoutput=1
\documentclass[%
 reprint,
 amsmath,amssymb,
 aps,prl,
]{revtex4-2}

\usepackage{graphicx}
\usepackage{dcolumn}
\usepackage{bm}

\begin{document}

\preprint{APS/123-QED}

\title{Extreme Lightwave Electron Field Emission  from a Nanotip}

\author{Dominique Matte}
 \affiliation{Centre for the Physics of Materials and Department of Physics, McGill University, Montreal, Quebec H3A 2T8, Canada}
\author{Nima Chamanara}
 \affiliation{Centre for the Physics of Materials and Department of Physics, McGill University, Montreal, Quebec H3A 2T8, Canada}
\author{Lauren Gingras}
 \affiliation{Centre for the Physics of Materials and Department of Physics, McGill University, Montreal, Quebec H3A 2T8, Canada}
\author{Laurent P. Ren\'{e} de Cotret}
 \affiliation{Centre for the Physics of Materials and Department of Physics, McGill University, Montreal, Quebec H3A 2T8, Canada}
\affiliation{Department of Chemistry, McGill University, Montreal, Canada}
\author{Tristan L. Britt}
 \affiliation{Centre for the Physics of Materials and Department of Physics, McGill University, Montreal, Quebec H3A 2T8, Canada}
  \affiliation{Department of Chemistry, McGill University, Montreal, Canada}
\author{Bradley J. Siwick}
 \affiliation{Centre for the Physics of Materials and Department of Physics, McGill University, Montreal, Quebec H3A 2T8, Canada}
 \affiliation{Department of Chemistry, McGill University, Montreal, Canada}
\author{David G. Cooke}
\email{cooke@physics.mcgill.ca}
 \affiliation{Centre for the Physics of Materials and Department of Physics, McGill University, Montreal, Quebec H3A 2T8, Canada}
 
\date{\today}

\begin{abstract}
We report on sub-cycle terahertz light-field emission of electrons from tungsten nanotips under extreme conditions corresponding to a Keldysh parameter $\gamma_K\approx10^{-4}$. Local peak THz fields up to 40~GV/m are achieved at the apex of an illuminated nanotip, causing sub-cycle cold-field electron emission and acceleration in the quasi-static field. By simultaneous measurement of the  electron bunch charge and energy distribution, we perform a quantitative test of quasi-static Fowler-Nordheim tunnelling theory under field conditions that completely suppress the tunnel barrier. Very high bunch charges of $\sim10^6$ electrons/pulse are observed, reaching maximum energies of 3.5~keV after acceleration in the local field. The energy distribution and emission current show good agreement with Fowler-Nordheim theory even in this extreme field regime. Extending this model to the single-shot regime under these conditions predicts peak electron distributions with a spectral purity of $10^{-4}$. THz field-induced reshaping and sharpening of the nanotip is observed, reducing the tip radius from 120~nm to 35~nm over roughly $10^9$ THz shots. These results indicate THz-driven nanotips in the extreme field limit are promising electron sources for ultrafast electron diffraction and microscopy. 
\end{abstract}

\maketitle

Light-matter interactions enter the extreme limit when the energy scale of the field interaction meets or exceeds a characteristic excitation energy of the system \cite{KrugerJPB2018,KrugerJPhysBAtomicMolOptPhys2012}. In electron photoemission from a metal surface, the relevant energy scale is the work function, typically a few~eV. The electron surface potential is transiently tilted in a light field applied normal to the surface, permitting electrons to tunnel to the vacuum at a rate that depends exponentially on the field-dependent width of the barrier \cite{FowlerRoySocLondon1928,MurphyPhysRev1956}. The ratio of the tunnelling time to the period of the light field defines the Keldysh parameter, $\gamma_K = \tau_{tun}/T$, and for $\gamma_K<<1$ the field interaction is quasi-static \cite{KeldyshJETP1965}. For near-infrared to visible light, the light intensities required to reach $\gamma_K = 1$ are on the order of $10^{13}$~W/cm$^2$, exceeding the damage threshold of metals \cite{WellershoffApplPhys1999}. Quasi-static light-matter interactions in the solid state have subsequently been relatively unexplored. 
\\
\indent Intense, single-cycle terahertz (THz) pulses illuminating metal nanotips provide a means to explore this extreme light-matter interaction regime \cite{WimmerNatPhys2014,HerinkNJP2014, LiNComms2016, EchternkampApplPhysBLasersOpt2016}. Localized to the tip apex, a broadband local field enhancement occurs via a lightning rod effect that scales as $\eta \sim\lambda/R$, where R is the nanotip radius and $\lambda$ is the vacuum wavelength \cite{LiNComms2016}. Intense single cycle THz pulses with peak electric fields of 1-10 MV/m become locally enhanced by $\eta > 1000$ to several GV/m at the nanotip surface. While such dc fields would immediately cause catastrophic damage, breakdown is strongly suppressed at higher frequencies under ac field conditions \cite{KilpatrickRevSciInst1957}. Moreover, under pulsed illumination the threshold field for breakdown scales with the pulse duration as $\tau^{-4}$\cite{WangProcIEEE1989}.  
\\
\indent These GV/m local fields exceed the threshold for sub-cycle cold-field emission from the metal nanotip \cite{HerinkNat2012,NanniNatComm2015, LiNComms2016,  RybkaNatPho2016}. Field-emitted electrons drawn directly from the Fermi level are subsequently accelerated to several keV in the quasi-static field before leaving the local field enhancement region that decays over length scales comparable to the tip radius. Thus the electrons follow a deterministic path governed by the near instantaneous field under which they are ejected. These cold-field emitted electrons have potential applications as seeds for THz-based accelerators and sources for time-resolved electron diffraction \cite{RopersPhysRevLett2007} and time-resolved point-projection electron microscopy \cite{QuinonezRevSciInst2013}. For many applications, particularly those seeking near single shot operation, electron bunch charges of $>10^6$ electrons/pulse are typically required for simple, few atom unit cells \cite{DaoudStrucDyn2017}. Electron bunch charges up to 100 electrons/pulse have been reported from nanotips driven by mid-infrared pulses with $\gamma_K\sim0.1$ \cite{HerinkNat2012}. We note that larger bunch charges were likely achieved, but not quantitatively measured, using low repetition rate, intense THz pulses \cite{LiNComms2016}. 
\begin{figure*}[t!]
\includegraphics[width=1\textwidth]{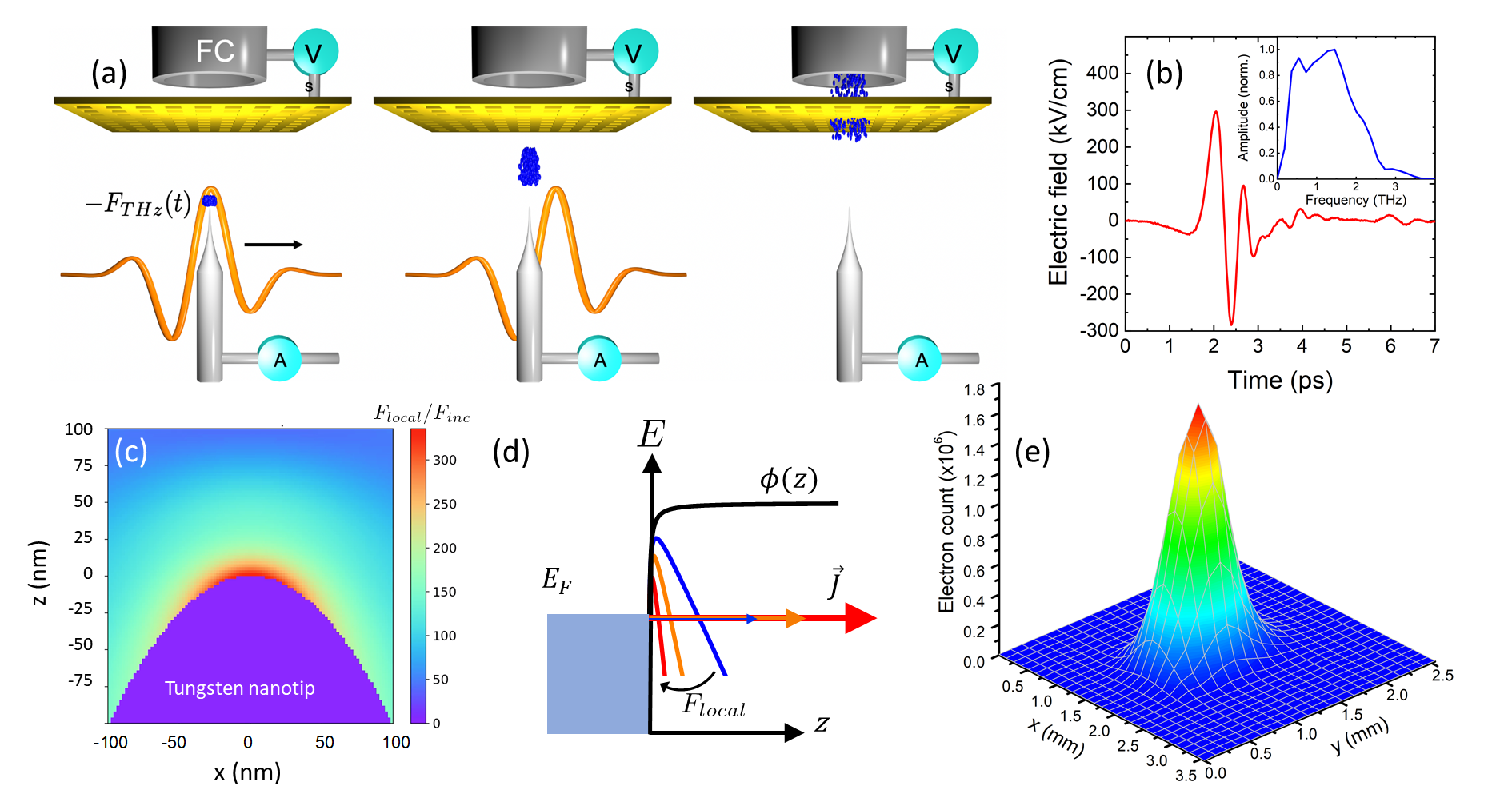}
\caption{(a) Schematic of the experiment, whereby a single cycle THz pulse, $F_{THz}(t)$, with polarization aligned along the nanotip axis illuminates a metal nanotip and ejects electrons towards a Faraday cup after passing through a grid held at a variable stopping potential $V_s$. (b) Electro-optic sampling of the incident THz pulses within the vacuum chamber illuminate the tip, with a peak THz field of 298 kV/cm and Fourier amplitude spectrum shown in the inset. (c) Three dimensional finite--difference time--domain simulation of the near-field enhancement normal to the tip apex in the vicinity of a 50 nm radius tip. (d) Schematic of the electron potential at the metal-vacuum interface and the field-induced tilting leading to a tunnel current. (e) The spatially resolved electron count per THz pulse as the nanotip is scanned through the fixed THz focal spot.}
\label{Fig1}
\end{figure*}
\\
\indent In this work, we demonstrate sub-cycle THz field emission of electron pulses from a tungsten nanotip in the extreme field limit corresponding to $\gamma_K\sim10^{-4}$, deep into the quasi-static regime. Average emission currents $>0.1$~nA are measured corresponding to bunch charges of $\sim10^6$ electrons/pulse. We simultaneously measure the electron energy distribution, with peak energies up to 3.5 keV, comparable to previous work \cite{LiNComms2016}. We test this emission against quasi-static Fowler--Nordheim theory and find good agreement despite it being well beyond the limitations of the model. Finally, a pronounced sharpening of the nanotip is observed under these extreme field conditions, gradually reducing the size of the tip from its initial 120 nm radius to sub-35 nm over an estimated $10^9$ shots.
\\
A schematic of the experiment is shown in Fig.~\ref{Fig1}(a). Single-cycle THz pulses are generated by tilted pulse-front optical rectification in a room temperature, MgO:LiNbO$_3$ prism pumped by 800 nm, 4 mJ laser pulses of 120 fs duration at a 1 kHz repetition rate \cite{YehAPL2007,HiroriAPL2011}. Intense THz pulses are focused to a near diffraction--limited spot on a tungsten nanotip with linear polarization aligned to the nanotip axis, held within a small vacuum chamber with a base pressure of $\sim10^{-7}$ torr. Tips were prepared from a polycrystalline tungsten wire using standard electrochemical etching in a KOH solution \cite{LucierThesis}. The nanotip radius was initially 120~nm, as verified by scanning electron microscopy (See Fig.~\ref{Fig4}).  The transmitted THz pulse is minimally perturbed by the presence of the nanotip and is 1:1 imaged on a 200~$\mu$m-thick, (110)-cut GaP crystal for free-space electro-optic sampling of the THz field. The most intense THz waveform experienced by the tip is shown in Fig.~\ref{Fig1}(b) with its amplitude spectrum shown in the inset. The fields are calibrated assuming a GaP electro-optic coefficient $r_{41} = 0.97$~pm/V \cite{yariv1975quantum}, accounting for the Fresnel loss of the final high resistivity silicon vacuum chamber window. A pair of wire grid polarizers was used to vary the incident THz peak field strength without changing its polarization state. The tungsten nanotip can be actuated in all three directions inside the chamber through the THz focus over a several mm range by moving the nanotip holder coupled through a baffle.

\indent The electron energy distribution and bunch charge are measured simultaneously via a Faraday cup detector positioned less than 1~cm away from the tip along the emission axis, with a varying stopping potential ($V_s$) applied to a grid placed before the detector. Additionally, an electrical connection is made to the tip allowing a direct measurement of the total current being drawn through the tip. These two currents are measured simultaneously using two Keithley 6517B electrometers. 
\begin{figure*}[th!]
\includegraphics[width=0.9\textwidth]{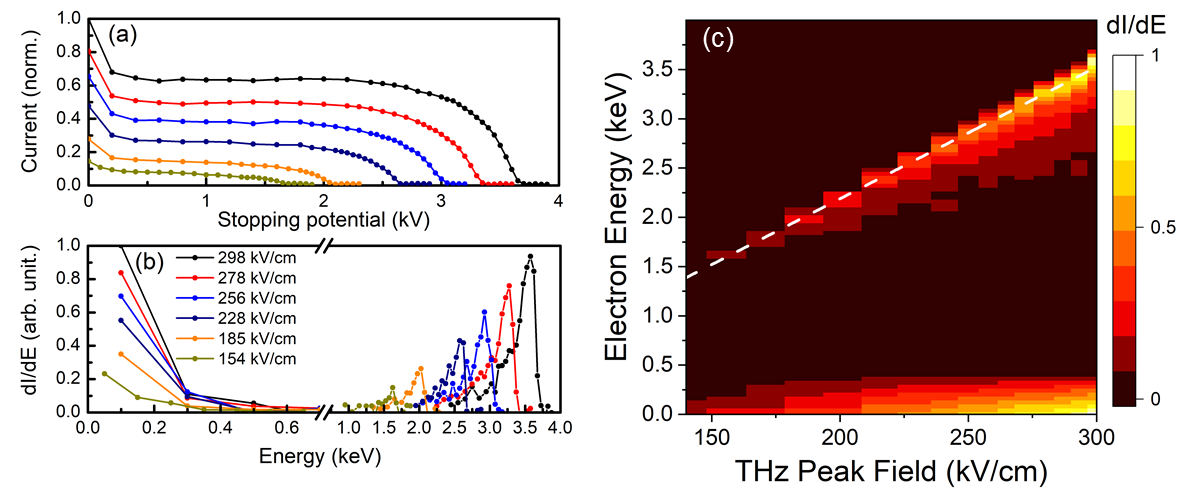}
\caption{(a) Electron emission current measured at the Faraday cup detector as a function of the applied retarding potential for THz peak field strength from 154 to 298 kV/cm.  Curves are normalized with respect to the 298 kV/cm max current. (b) Energy distributions derived from the raw data, which show low and high energy peaks respectively which correspond to two qualitatively distinct types of electrons emission. Low energy tail in the high energy peak were cut out for clarity. (c) Mapping of the energy distribution as a function of the THz peak field strength.}
\label{Fig2}
\end{figure*}

\indent Three dimensional finite--difference time--domain (FDTD) simulations were performed to model the local THz fields coupled to tungsten nanotips. The nanotip shape was approximated as a prolate spheroid, which allowed the expansion of the fields in spherical coordinates where the Helmholtz equation is separable \cite{SinhaRS1977}. Typical field enhancements of $\sim350$ are calculated for a 50 nm radius tip, as shown in Fig.~\ref{Fig1}(c) using the input field given in Fig.~\ref{Fig1}(b) (see Supplemental Material \cite{Supplemental}). While this dipole model qualitatively captures the decay of the local field enhancement away from the tip apex, it ignores the atomic scale structure of the tip and spatial texture in the local field that would be observed in a field ion microscopy image, for example. A more accurate estimation of the local field at the point of emission can be obtained from the measured energy distribution as discussed below.
\\
\indent Cold field emission of electrons from metal nanotips under dc and quasi-dc field conditions, depicted in Fig.~\ref{Fig1}(d), is typically described by Fowler-Nordheim theory \cite{FowlerRoySocLondon1928}. The instantaneous local THz field, $F_{loc}$, tilts the electron potential given by the Schottky-Nordheim function $V(z) = \phi - e F_{local}z - \frac{e^2}{16\pi \epsilon_0 z}$
with $\phi = 4.5$~eV being the work function of the tungsten tip. The critical field required to lower this potential barrier to zero relative to the Fermi energy is given by $F_\phi = \frac{4\pi\epsilon_0\phi^2}{e} = 14$~GV/m for tungsten \cite{ForbesJVSTB2008}. The emission current $\vec{J}$ is subsequently calculated via the tunneling probability through this barrier. The zero temperature Fowler-Nordheim equation for the current density J can then be written as
\begin{equation}
  J(t)=\frac{a F_{loc}(t)^2}{F_\phi^2\phi}\exp\biggl[-v b \frac{F_\phi}{F_{loc}(t)}\biggr]
\end{equation}
where a and b are Fowler-Nordheim constants. The function $v\approx 1 - F_{loc}/F_\phi$ accounts for image charges and exponentially suppresses the emission, although is valid only for $F_{loc}<F_\phi$  \cite{ForbesJVSTB2010}. Since the influence of the image charges is limited to very close to the tip apex where local fields are the strongest, we find all data can be described by setting $v = 1$ (see Supplemental Material \cite{Supplemental}).
\\
\indent The total electron emission ($V_s = 0$~V) is shown in Fig.~\ref{Fig1}(e), measured using a direct electrical connection on the nanotip and scanning the tip through the focal plane of the THz pulse.  Electron bunch charges in excess of $10^6$ electrons/pulse were measured on peak, more than three orders higher than previously reported \cite{BormannPRL2010,HerinkNat2012}. The relative THz/tip-position dependence of the field-emission current (bunch charge) follows an asymmetric Gaussian with FWHM of 580~$\mu$m and 450~$\mu$m in the x and y directions, respectively (Fig. 1(e)).  These dimensions are comparable to those of the THz intensity distribution at the tip position as measured using a microbolometer camera. While Fowler-Nordheim emission predicts a sharpened distribution for $F_{loc}<<F_\phi$ due to the exponential dependence on the field, in the extreme limit of $F_{loc}>>F_\phi$ the exponential term saturates and the electron distribution is governed by $F_{loc}^2$, i.e. the intensity distribution. Thus both the large electron count rate and the Gaussian spatial distribution indicate that we are operating in the extreme limit of field emission.
\\
The normalized electron current from the Faraday cup is shown in Fig.~\ref{Fig2}(a) for varying peak incident THz field strength. The current shows a marked reduction at low potentials ($<500$ eV) and a high energy cutoff at several keV. We note the beam current could be manipulated by the presence of a permanent magnet brought close to the chamber, confirming the origin of these signals as a free-space electron beam. The numerical derivative of this spectra gives the energy distribution, whose low and high energy regions are shown in Figs.~\ref{Fig2}(b). The low energy distribution shows a monotonically decreasing distribution to a cutoff of approximately 400 eV, while the high energy distribution is sharply peaked at energies up to 3.5 keV. The origin of the low energy distribution is puzzling. Given the significant portion of electrons present in the low energy part of the distribution, these electrons must be field emitted during the peak of the THz field. Electrons emitted during this time would usually be accelerated ballistically to form the high energy peaked distribution. Evidently, a significant portion of electrons are not immediately swept out of the tip region but instead experience a lower accelerating THz field. Such a delayed photoemission channel has recently been demonstrated using few-cycle optical pulses, where re-scattered electrons are driven back to the nanotip surface by subsequent field cycles \cite{YanagisawaSciRep2016}. Inelastic scattering within the metal can result in delay of the electron emission by 10's of fs. While this cannot explain our results, it points to the role of inelastic scattering during the emission process. A possible mechanism for such a significant low energy population is through interactions with surface contaminants. Initially emitted electrons can scatter within contaminant layers \cite{TanumaSurfIntAnalysis2003}, subsequently becoming trapped. They can eventually escape and accelerate in a subsequent cycle of the THz field. Numerical simulations of THz resonances excited in the nanotip shows that a radially polarized, dispersionless Sommerfeld wave is launched on the nanotip \cite{WangNat2004, WangJOptSocAmB2005}. This wave reflects at the tip boundaries resulting in a multi-cycle field at a resonant frequency of $\omega_0=(2\pi)c/2L$ where $L\sim5$~mm is the tip length. As the local field enhancement favours low frequencies (numerically $\gamma\propto\omega^{-1.4}$, see Supplemental Material \cite{Supplemental}), as does the ponderomotive energy $U_p = e^2F_{loc}^2/4 m\omega_0^2$, electrons can reach 400~eV in this multi-cycle field with $F_{loc}$ as low as $35$~MV/m. The complete energy distribution is shown in Fig.~2(c). While the ponderomotive energy of trapped electrons should scale as $F_{loc}^2$, we observe saturation in the peak electron energy at $\sim400$~eV. While the electron energy loss spectra of tungsten nitride shows no excitations in the 100-400 eV energy range, an onset of strong electron absorption above 400 eV coincides with the nitrogen 1s excitation \cite{SotoASS2003,SunTSF2004,WicherVacuum2019}. We therefore assign the low energy distribution to electrons inelastically scattered within the nitride, delaying emission and accelerated by the Sommerfeld resonance of the tip. Further evidence for electron energy transfer to surface adsorbates is given later when we examine the nanoscale structure of the tip after long term exposure.
\\
\indent A linear field dependence of the high energy peak is observed in Fig.~\ref{Fig2}(c), as expected for ballistic acceleration. The peak energy is given by $U_{max} = e F_{loc}l_F$ where $l_F$ is the effective length scale of the field enhancement. The linear fit (white dashed line) with a slope of 13.3~eV~cm/kV = $e\gamma l_F$ directly yields the local field enhancement factor $\gamma = 3800$ for a $l_F = 35$ nm, approximated as the tip radius. The $F_{loc}$ in this simple estimate is as high as 115 V/nm, more than 8 times the critical field $F_{\phi}$. The time scales for electrons to leave the field enhancement region is on the order of 5 fs, well within the quasi-static emission regime. The corresponding adiabaticity parameter $\delta = l_F m \omega^2/eF_{local}\approx 0.001$, relating the near field decay length to the electron quiver amplitude, is also well within the quasi-static limit \cite{HerinkNat2012}.

\begin{figure}[t!]
\includegraphics[width=1.0 \columnwidth]{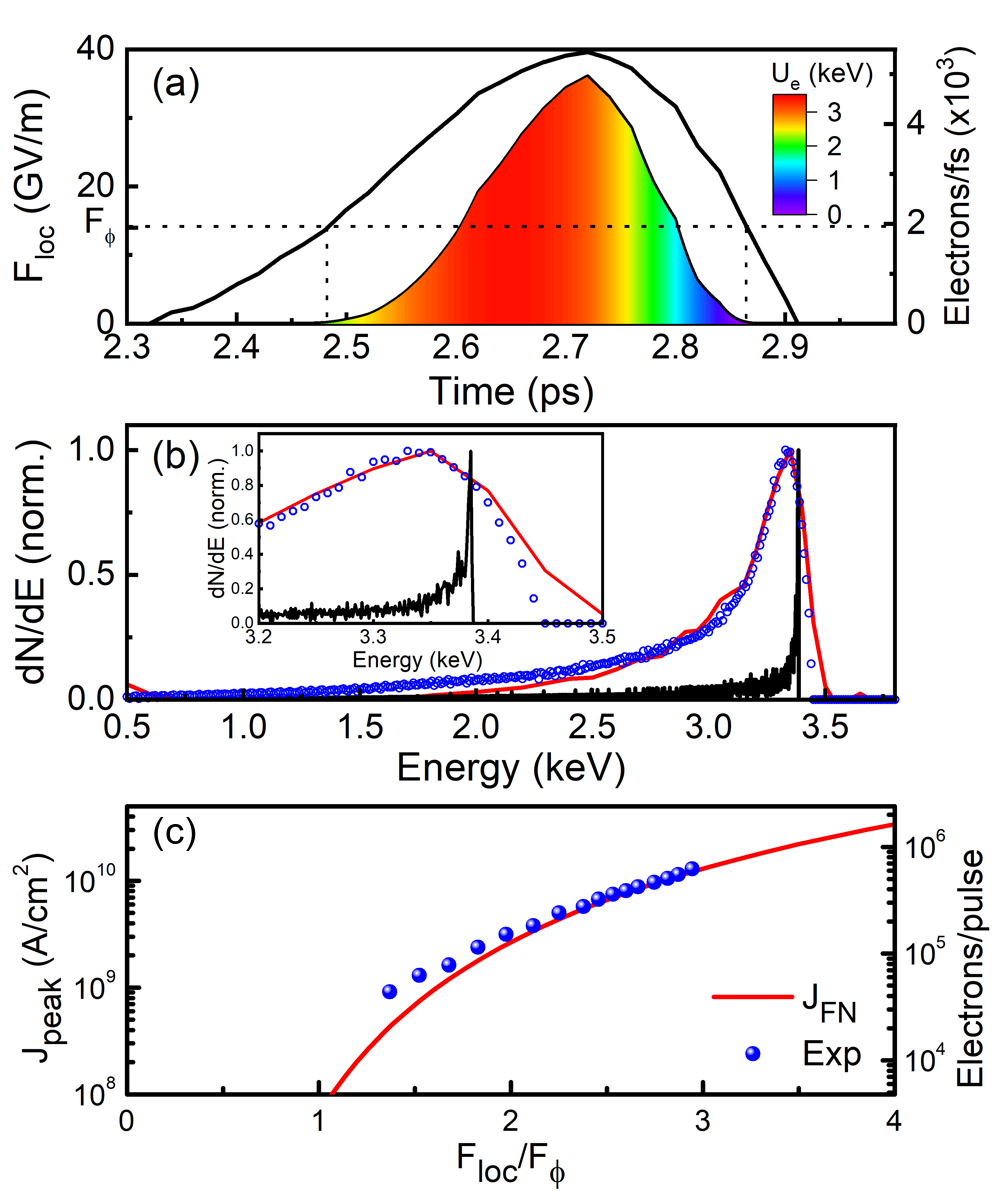}
\caption{(a) Local THz pulse electric field F$_{loc}$ and simulated electron emission rate as described in the text. The chirp of the electron pulse is shown in the color scale. (b) Experimental normalized electron energy distribution (blue circles) at an incident THz peak electric field of 285kV/cm. A simulation of the single shot normalized electrons energy distribution (black line) using Fowler-Nordheim cold-field emission theory and ballistic acceleration of the electrons by the local electric field, as well as with a 2\% Gaussian peak field fluctuation (red line). (inset) Data plotted in the high-energy portion of the energy spectrum. (c) THz peak field dependence of the total high energy electron per pulse from the nanotip. Low energy electrons (below 600~eV) are removed from the electron yield.}
\label{Fig3}
\end{figure}

In the quasi-static limit, the Fowler-Nordheim (F-N) equations can simulate the emission dynamics of the high energy distribution. The emission current and final electron energy were calculated using a time-dependent, one dimensional finite-element simulation taking into account the near-field decay of a hyperboloidal tip  $\eta/[1~+~2x/R]$, \cite{LiNComms2016,Miller1996atom} with $\eta=1380$ and R=35~nm, the Gaussian THz field focus and the THz pulse waveform in Fig.~\ref{Fig1}(b). The high energy electrons rapidly leave the field enhancement region, however still experience the entire THz waveform before leaving the THz focal region. Electrons are only slightly slowed down by the subsequent field half-cycle before they escape the free-space focal spot. Fig.~\ref{Fig3}(a) shows the local field half-cycle and the resultant electron emission rate up to several thousand electrons/femtosecond and occurring in a sub-cycle burst approximately 200 fs in duration. The onset of such large emission rates occurs when the local field reduces the potential barrier to zero, or $F_{loc} = F_\phi$ (see Supplemental Material \cite{Supplemental}). Electrons are chirped in energy according to their near-instantaneous acceleration through the local field in varying parts of the cycle, with the highest energies being emitted in a sub-100 fs duration. At these field strengths, F-N theory predicts a much sharper peaked energy distribution than we and others have observed experimentally \cite{HerinkNJP2014,LiNComms2016}, shown in Fig.~\ref{Fig3}(b) as a black line with a FWHM of $\approx$1~eV on the high energy peak.  To simulate experimental conditions that average over thousands of laser shots, we add statistical fluctuations of the THz pulse amplitude represented by a truncated Gaussian distribution with a standard deviation of 2~\% centered on the measured THz peak field, a good approximation to our shot-to-shot fluctuations. The calculated energy distribution is obtained by averaging and is shown in Fig.~\ref{Fig3}(b) (red line), in excellent agreement with the measured high energy distribution (blue circles). Moreover, the inset shows the single-shot distribution need not match the experimentally observed emission peak energy due to the asymmetric emission process and chirp under the half-cycle of the pulse. Thus the single-shot energy distribution is expected to be extremely monochromatic in this high field regime, with a spectral purity on the order of $10^{-4}$. Thus the THz driven nanotip is expected to be an extremely bright electron source, potentially useful for single shot ultrafast electron microscopy and diffraction experiments with sub-100 fs temporal resolution.
\\
\indent The peak current density $J_{peak}$ and bunch charge/pulse under varying THz peak field is shown in Fig.~\ref{Fig3}(c). At the maximum fields, $J_{peak}$ exceeds $10^{10}$~A/cm$^2$ with the maximum bunch charge of $10^6$ electrons/pulse. Such current densities under dc field conditions would be completely dominated by space-charge effects, governed by the Child-Langmuir Law with $J\propto (F_{loc}/F_\phi)^{3/2}$ \cite{BarbourPhysRev1953}. The onset of space-charge effects is expected to occur for $F_{loc}/F_\phi\sim 0.5$ for dc field emission \cite{ForbesJVSTB2008}. For pulsed field emission from a nanotip, however, space-charge effects should be suppressed under the condition that the pulse duration ($\sim200$~fs) is much longer than the electron transit time through $l_F$ ($<5$~fs). We find that in this extreme strong-coupling regime, space-charge effects are negligible as the emission is well described by a simple one-dimensional Fowler-Nordheim theory with no compensation for local image charges or screening ($v=1$).

\begin{figure}
\includegraphics[width=0.7 \columnwidth]{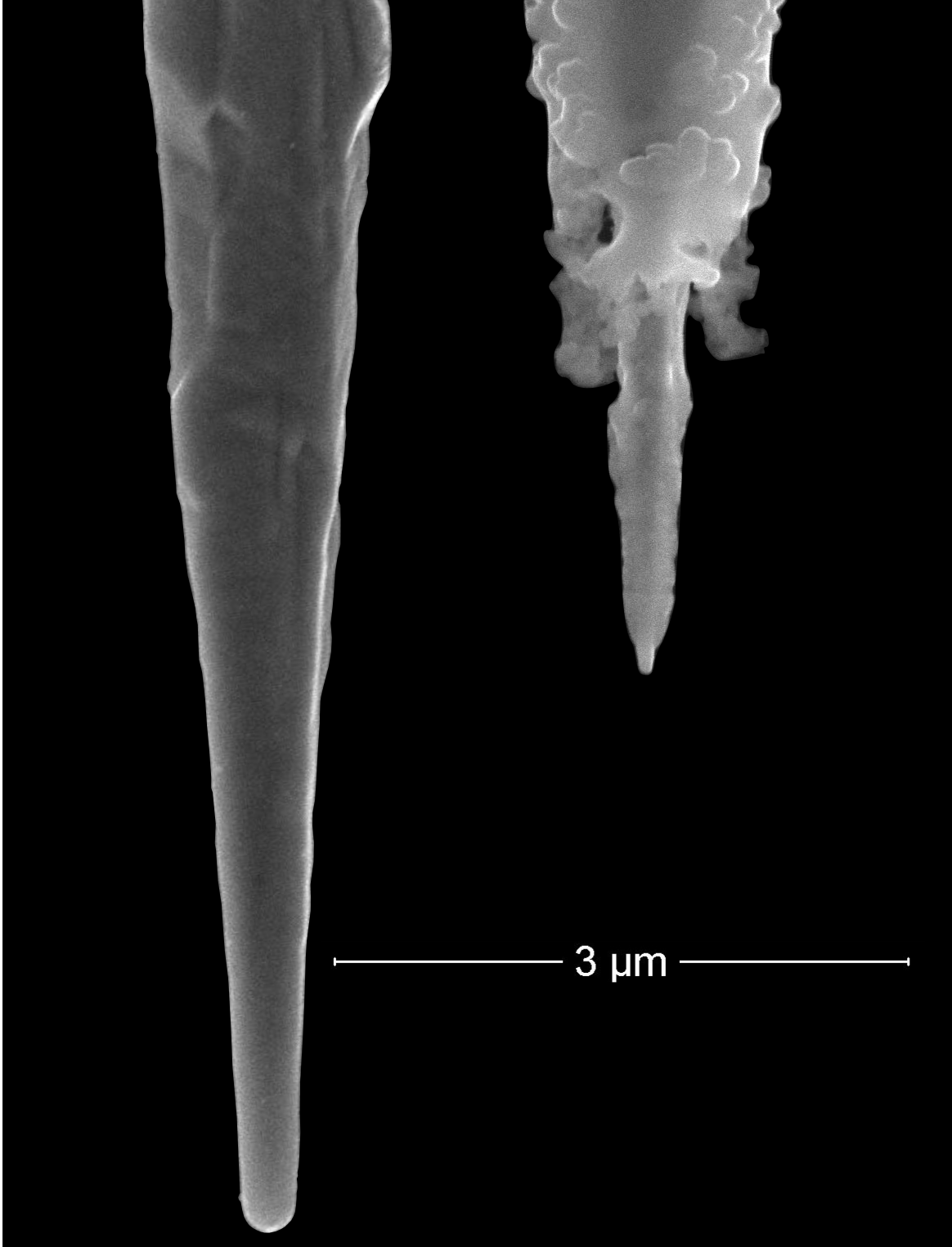}
\caption{ SEM image of a tungsten nanotip without exposure (120~nm radius)(left) and with exposure to $\sim1$~billion of THz shot (35~nm radius)(right) showing a strong reshaping/sharpening of the apex of the tungsten nanotip improving the electrons emission rate and energy.}
\label{Fig4}
\end{figure}

Comparative scanning electron microscope images of a freshly fabricated tungsten nanotip and a tip exposed to $\sim10^9$ high field THz pulses are shown in Fig.~\ref{Fig4}, showing a strong reshaping of the apex of the tip. Over the course of several weeks of experiments, the tip length was reduced by approximately 3~$\mu$m in length and the tip apex sharpened from a $\sim120$~nm to $35$~nm radius. We note that over the course of an experiment in several hours to days, the emission rate was stable. Such field-induced sharpening of a nanotip has been previously observed in field-ion microscopy under applied dc bias voltage, and was used to produce single atom-terminated nanotips \cite{RezeqJChemPhys2006}. The mechanism for such dc field reshaping was determined to be the formation of tungsten nitride, causing a protrusion that in turn caused a local field enhancement. The local field then exceeded the evaporation threshold for the nitride and the atoms were removed. We therefore tentatively attribute the observed sharpening of the nanotip under intense THz fields to field evaporation of tungsten nitride on a shot-to-shot basis. However given the observed low energy electron distribution, we consider an alternative mechanism could be inelastic scattering of accelerated electrons within the nitride causing the removal of atoms in a similar manner. In this case, electrons transfer energy to regions with tungsten nitride, which are then selectively evaporated in the field thus sharpening the tip over time on a shot-to-shot basis.

In conclusion, we have demonstrated sub-cycle THz driven cold-field electron emission from a tungsten nanotip under extreme local field conditions. Electron bunch charges and peak current densities on the order of $10^6$ electrons/pulse and $10^{10}$ A/cm$^2$ are quantitatively measured. Energies distribution up to 3.5~keV are observed and are accurately described by Fowler-Nordheim emission with ballistic acceleration in a quasi-static field. Seeing no signature of saturation in the emission current for increasing $F_{loc}$, combined with the field-induced sharpening of the nanotip and the predicted single-shot spectral purity of $10^{-4}$ indicates that THz driven nanotip field emission holds promise for ultra-bright, ultrafast electron sources.

\section*{Acknowledgements}{D. G. C. and B. J. S.  gratefully acknowledge support from FRQNT, NSERC and the Canadian Foundation for Innovation (CFI). We thank Simon L. Lange and Prof. Peter Uhd Jepsen (Technical University of Denmark) for initial discussions and 3D simulations. We also thank Prof. Frank Hegmann (University of Alberta) for useful discussions. 
}

\bibliography{ElectronEmission}

\end{document}


\preprint{AIP/123-QED}

\title{\centering{SUPPLEMENTAL MATERIAL\\Extreme Lightwave Field Emission of Electrons from a Nanotip}}

\author{Dominique Matte}
 \affiliation{Centre for the Physics of Materials and Department of Physics, McGill University, Montreal, Quebec H3A 2T8, Canada}
\author{Nima Chamanara}
 \affiliation{Centre for the Physics of Materials and Department of Physics, McGill University, Montreal, Quebec H3A 2T8, Canada}
\author{Lauren Gingras}
 \affiliation{Centre for the Physics of Materials and Department of Physics, McGill University, Montreal, Quebec H3A 2T8, Canada}
\author{Laurent P. Ren\'{e} de Cotret}
 \affiliation{Centre for the Physics of Materials and Department of Physics, McGill University, Montreal, Quebec H3A 2T8, Canada}
\affiliation{Department of Chemistry, McGill University, Montreal, Canada}
\author{Tristan L. Britt}
 \affiliation{Centre for the Physics of Materials and Department of Physics, McGill University, Montreal, Quebec H3A 2T8, Canada}
  \affiliation{Department of Chemistry, McGill University, Montreal, Canada}
\author{Bradley J. Siwick}
 \affiliation{Centre for the Physics of Materials and Department of Physics, McGill University, Montreal, Quebec H3A 2T8, Canada}
 \affiliation{Department of Chemistry, McGill University, Montreal, Canada}
\author{David G. Cooke}
\email{cooke@physics.mcgill.ca}
 \affiliation{Centre for the Physics of Materials and Department of Physics, McGill University, Montreal, Quebec H3A 2T8, Canada}

\date{\today}
\maketitle
\onecolumngrid


\beginsupplement

\section{\label{sec:level1} Time-dependent Finite Element Simulations - Electron energy distribution}

Time-dependent finite element simulations have been performed to simulate the electron emission during the THz pulse to fit the electron energy distribution using the field enhancement factor as the only fitting parameter. 

\subsection{Quiver to sub-cycle emission regime}
 
\begin{figure}[h]
    \centering
    \includegraphics[width=0.7\columnwidth]{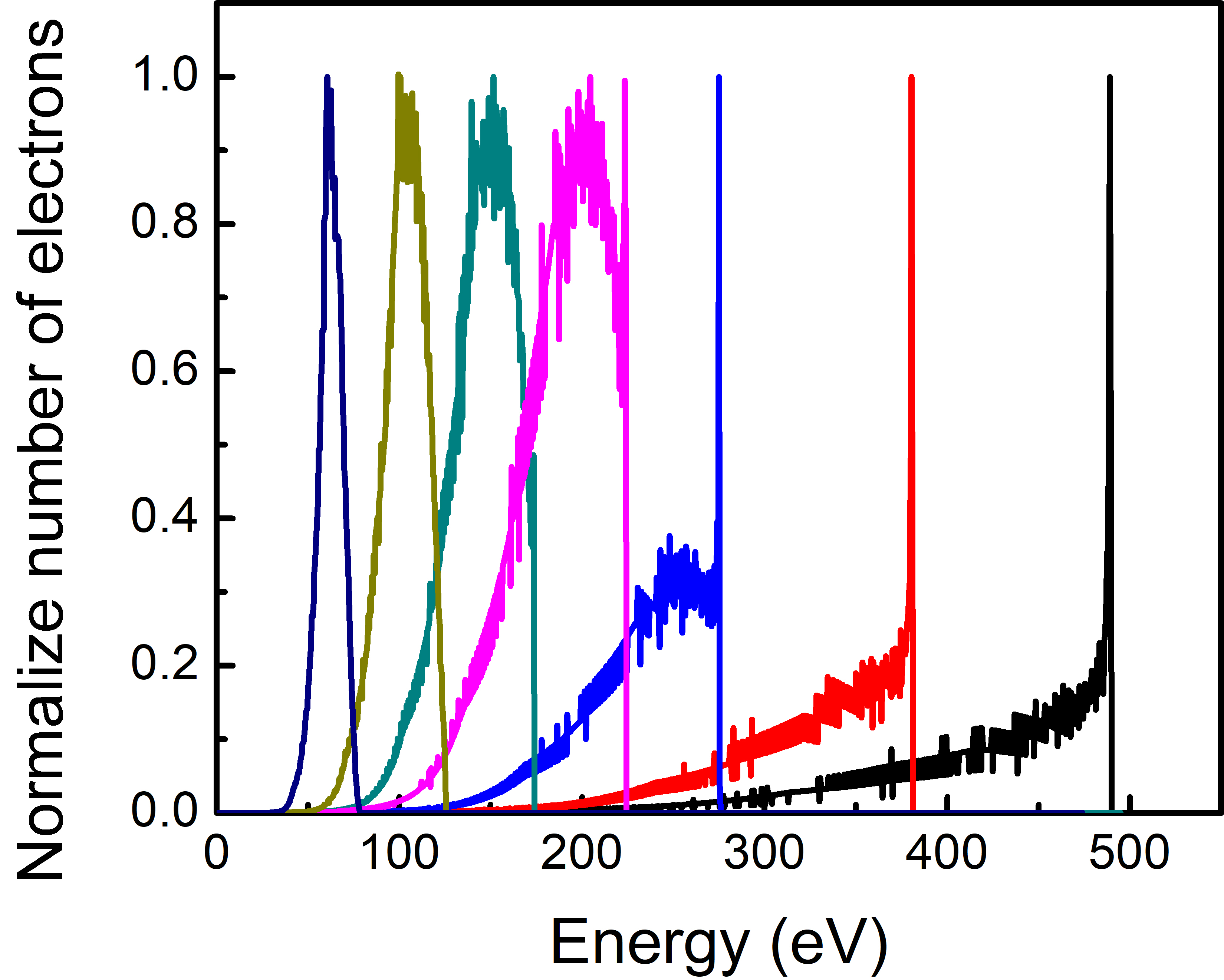}
    \caption{Time-dependent finite element simulations of the normalized electron energy distributions for peak THz electric fields of 10, 15, 20, 25, 30, 40 and 50~kV/cm, considering a field enhancement at the tip apex of 1380.}
    \label{FigS1}
\end{figure}

 In light-field electron acceleration, two regimes are relevant: the quiver regime and the sub-cycle regime \cite{HerinkNJP2014}. The sub-cycle, quasi-static regime is characterised by a sharp high energy peak in the electron energy distribution, tailing to lower energies. Electrons are ejected at the highest emission rate near the peak of the THz lightwave, and are accelerated out of the field enhancement region on time scales much shorter than the cycle time. Their final energy is then directly related to the field at the time of their emission. At lower fields, electrons emitted during the THz peak do not leave the enhancement region fast enough and are slowed down by the next half-cycle or even multiple cycles of the THz pulse, leading to a broadened peak. This multi-cycle interaction is the so-called quiver regime. Simulations of the emission process were performed to find the transition between the two regimes under our experimental conditions. We input our experimentally measured THz electric field waveforms obtained by free-space electro-sampling as our incoming light field. A field enhancement factor of $\gamma$=1380 was found to describe the energy distribution with the functional form of the near-field decay taken to be that of a hyperboloidal tip $\frac{\gamma}{1+\frac{2x}{R}}$, \cite{LiNComms2016,Miller1996atom} for a R=35~nm tip radius. We used the zero-temperature Fowler-Nordheim equations (with $v=1$) to obtain the electron emission at any given time during the THz pulse in the quasi-static limit (Keldysh parameter $\gamma_K<<1$). The energies of the electrons are calculated using a time-dependent finite element simulation including the THz pulse Gaussian focus. For our experimental THz peak fields, between 140~kV/cm and 300~kV/cm, all simulations showed characteristic sub-cycle emission. Figure \ref{FigS1} shows the normalize energy distribution for THz peak fields of 10 to 50~kV/cm. A clear transition between the quiver regime and the sub-cycle regime is exhibited. The high energy peak appears around 25~kV/cm and becomes dominant at 30~kV/cm. Already at 50~kV/cm, the distribution resembles the 285~kV/cm energy distribution shown in Fig.~3 of the main manuscript. It is important to note that because the pulsed nature of the experiment and the exponential behavior of the field emission equation, the emission rate per pulse is greatly reduced at lower THz fields with rates below 1 electron per pulse at 40~kV/cm. We were unable to measure any emission experimentally with our current setup to go through the transition regime. 
 
\subsection{Critical field (F$_\phi$)}
\begin{figure}
    \centering
    \includegraphics[width=0.7\columnwidth]{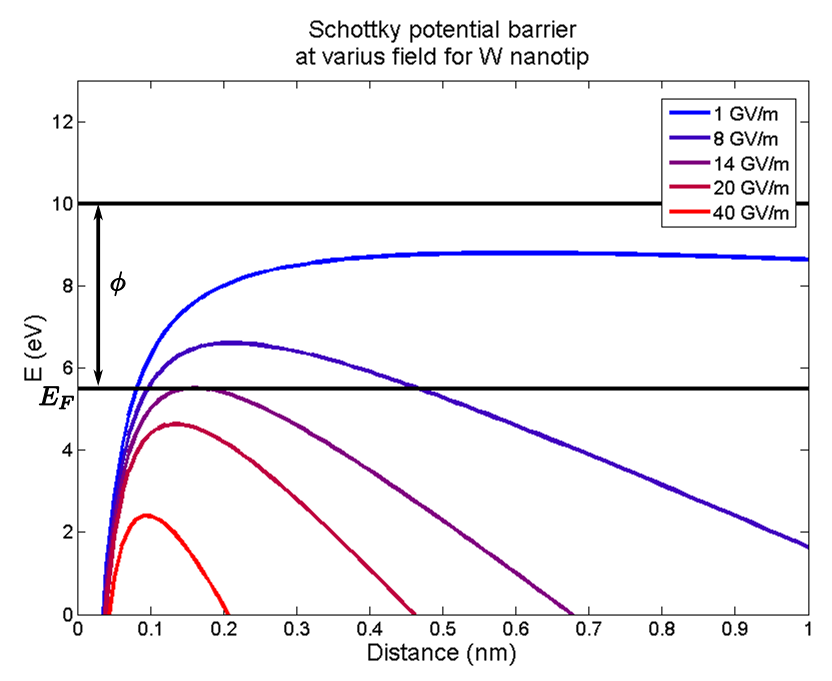}
    \caption{Schottky-Nordheim potential barrier for electron cold field emission at different local field strength. }
    \label{FigS2}
\end{figure}

Cold field emission occurs when a high electric field is applied to a material and bends the potential barrier enough that electrons can start tunnelling from an occupied energy level within the material. The Schottky-Nordheim potential is given by the work function $\phi = 4.5$~eV for tungsten minus the potential of the local electric field and the image potential created by the electrons leaving the material, 
$V(z) = \phi - e F_{local}z - \frac{e^2}{16\pi \epsilon_0 z}$.
Fowler-Nordheim theory assumes that the emitted electrons come from the Fermi energy. For high enough field values, the height of the potential barrier can be reduce to zero at the critical field, $F_\phi = \frac{4\pi\epsilon_0\phi^2}{e} = 14$~GV/m, and can even go below the Fermi energy. Figure \ref{FigS2} shows the potential barrier of a tungsten tip ($\phi~=~4.5~eV$), for local electric field below, above and at the critical field. Above the critical field, electrons at the Fermi energy do not have any barrier to tunnel through making the theory inapplicable. Thus we elected to use a simpler version of the Fowler-Nordheim equation by removing the image term in the potential and consider a simple triangular barrier of height $\phi$ above the Fermi energy. 

\begin{figure}
    \centering
    \includegraphics[width=0.7\columnwidth]{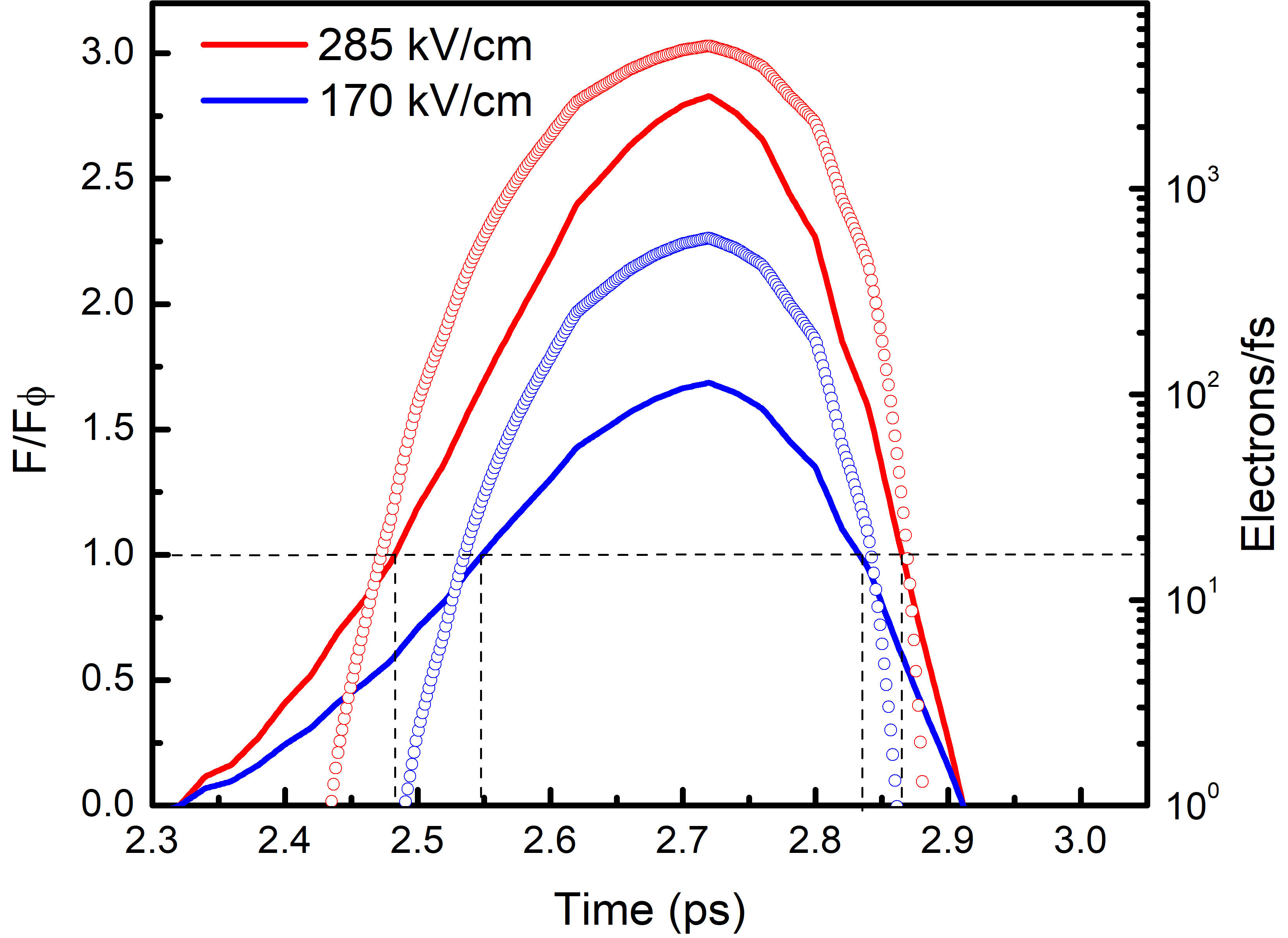}
    \caption{The normalized local electric field (line) and the electrons emission rate (open circles) for a THz peak field of 285~kV/cm (red) and 170~kV/cm (blue). }
    \label{FigS3}
\end{figure}

The local electric field used in our experiment exceeds the critical field (F$_\phi$). Figure \ref{FigS3} shows that the onset of the emission occur at that critical field for both 285~kV/cm and 170~kV/cm THz peak electric field. Also, the exponential term in the equation is less important at high field and the electron pulse duration is reduced at lower fields. 

\section{Three dimensional finite difference time domain (FDTD) simulations - Field enhancement}

Numerical three-dimensional finite-difference time domain simulations have been performed to gain insight on the local field enhancement and its decay, modelling a tungsten nanotip as a prolate spheroid. The measured THz transient was used as the input field with the light polarization aligned with the tip axis.

\begin{figure}
    \centering
    \includegraphics[width=0.9\columnwidth]{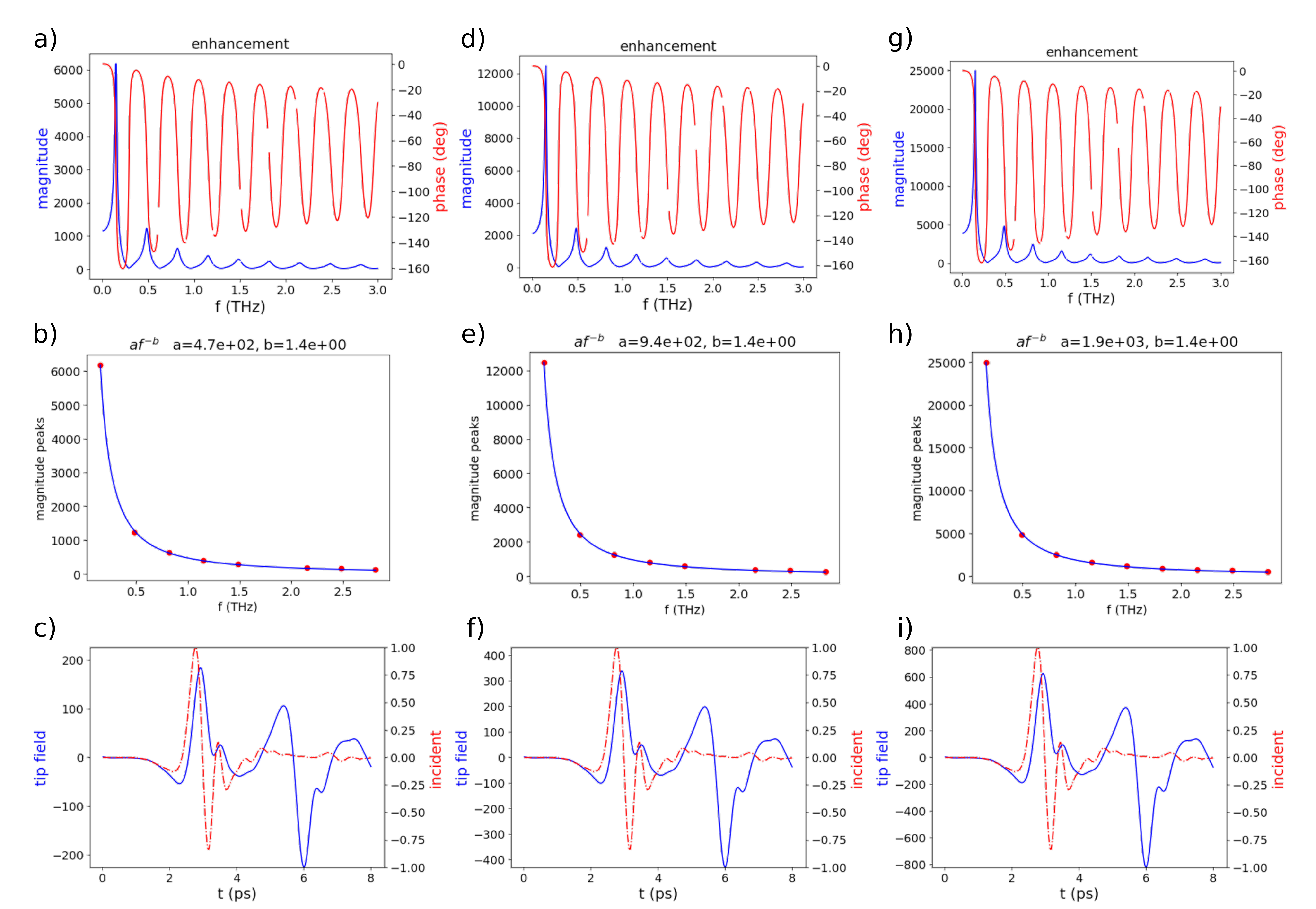}
    \caption{Three-dimensional finite-difference time domain (FDTD) simulations calculating (a, d, g) the magnitude (blue) and phase (red) of the enhancement factor with respect to the field frequency. (b, e, h) Fit of the decay in the peaks of the enhancement factor magnitude. (c, f, i) Comparison of the incident electric field normalize by the experimental THz field (red) and the reconstruct field at the apex of the tip (blue). Calculations have been made at three different tip radius: R=100~nm (a,b,c), R=50~nm (d,e,f) and R=25~nm (g,h,i). A nanotip length of 0.9~mm was use in all calculations.}
    \label{FigS4}
\end{figure}

 Simulations were performed at different tip radii: R=25~nm, R=40~nm and R=100~nm for a tip length L=0.9~mm (Fig.\ref{FigS4}). The magnitude of the field enhancement factor exhibits a periodic frequency (f) dependence shown in Fig.~\ref{FigS4} (a), (d) and (g), reflecting the macroscopic resonances of the tip at frequencies $\omega_0=mc/2L$ where $m$ is an integer. Computational run times limit the length of the tip to less than 1-2 mm. We note the free spectral range for the resonances for our experimental setup (tip length = 8 mm) is only 20 GHz, so narrow to have the effect of a continuous broadband enhancement. The peaks of the periodic enhancement factor follow a frequency dependence of $\eta\propto\omega^{-1.4}$ shown in Figs.~\ref{FigS2} (d), (e) and (h). Figure \ref{FigS4} (c), (f) and (i) compare the normalize incident field (red) and the effective electric field at the tip (blue). The enhancement factor vary from around 180 to 620 following a $\gamma=aR^{b}$ with $b\approx-0.9$. 
 
\begin{figure}
    \centering
    \includegraphics[width=0.7\columnwidth]{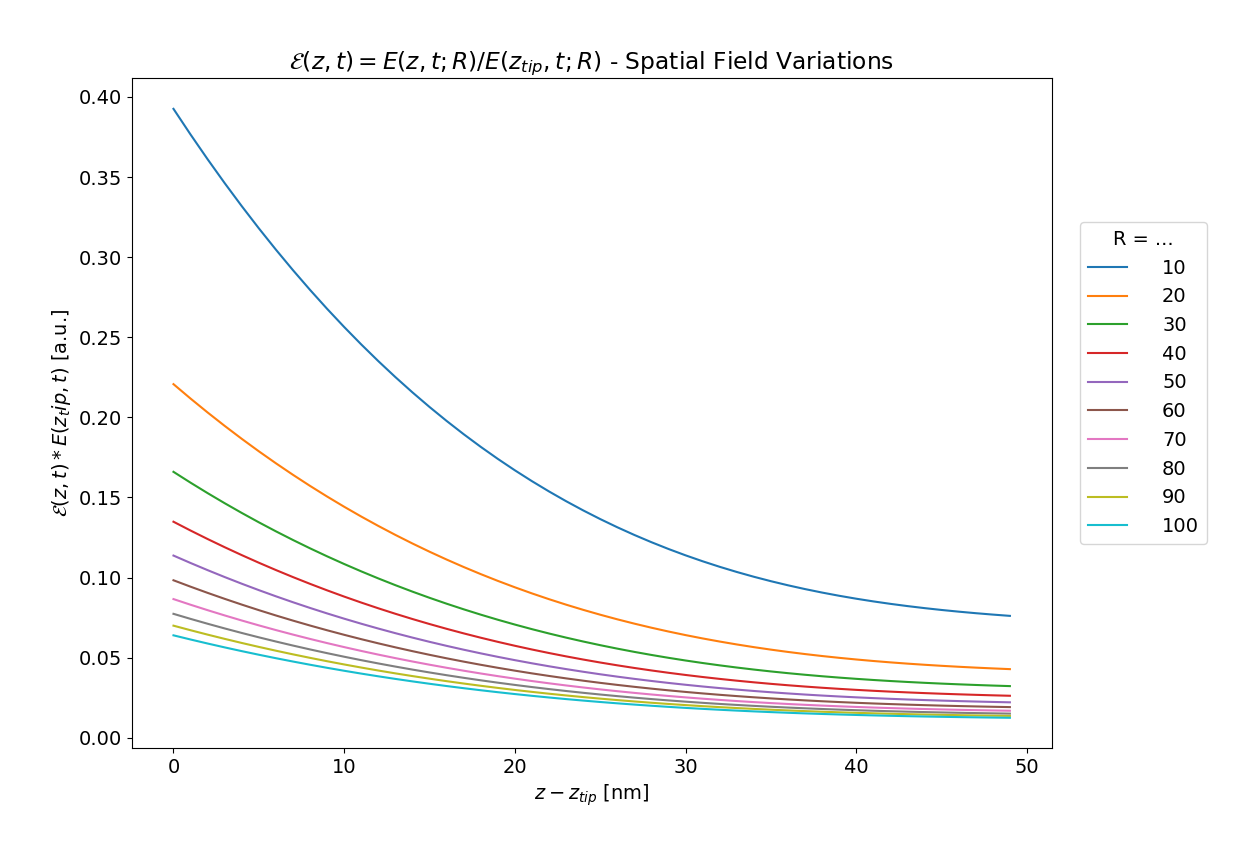}
    \caption{Field enhancement decay from the apex of the tip for tip radius from R=10~nm to R=100~nm.}
    \label{FigS5}
\end{figure}

The field enhancement decay is obtained by taking a cut of the field along the tip axis at the peak of the THz pulse (Fig. \ref{FigS5}). The decay length is found to be independent of the tip radius and only the magnitude increases when the tip radius decreases. The best fitting function for the decay is a bi-exponential. Given that the simulation does not seem to capture the physics of the experiment with both a much smaller field enhancement and lower energy distribution compare to the experiment, we used a hyperbolic decay for the energy distribution calculation following previous work \cite{LiNComms2016,Miller1996atom}.








\bibliography{ElectronEmission}